\documentclass[showpacs,preprintnumbers,amsmath,amssymb]{revtex4}
\usepackage{graphicx}
\usepackage[dvips]{epsfig}
\usepackage{dcolumn}
\usepackage{bm}
\def \be{\begin{equation}}
\def \ee{\end{equation}}
\def \bdm{\begin{eqnarray}}
\def \edm{\end{eqnarray}}
\begin{document}
\title{Forms of Eulerian correlation functions in the solar wind}
\author{A. Shalchi}
\affiliation{Institut f\"ur Theoretische Physik, Lehrstuhl IV:
Weltraum- und Astrophysik, Ruhr-Universit\"at Bochum,
D-44780 Bochum, Germany}
\date{\today}
\begin{abstract}
Current spacecraft missions such as Wind and ACE can be used to determine magnetic correlation 
functions in the solar wind. Data sets from these missions can, in principle, also be used to compute 
so-called Eulerian correlation functions. These temporal correlations are essential for understanding
the dynamics of solar wind turbulence. In the current article we calculate these dynamical correlations 
by using well-established methods. These results are very useful for a comparison with Eulerian
correlations obtained from space craft missions.
\end{abstract}
\maketitle
\section{Introduction}
In the past idealized models for turbulent fluctuations which can be found in the solar wind plasma 
or in the interstellar medium have been proposed (e.g. Matthaeus et al. 1995). We are concerned 
with statistically axisymmetric models of magnetostatic fluctuations $\delta \vec{B} (\vec{x})$ that 
are transverse to a uniform mean magnetic field $\vec{B}_0$. If solar wind turbulence is considered,
the mean field might be identified with the magnetic field of the Sun. The total magnetic field is a 
superposition of this mean field and the fluctuations $\vec{B}(\vec{x})=\vec{B}_0+\delta \vec{B}(\vec{x})$. 
Whereas we usually approximate the mean field by a constant field aligned parallel to the $z-$axis 
($\vec{B}_0 = B_0 \vec{e}_z$), the turbulent contribution has to be replaced by turbulence models. 
Some prominant examples are slab, 2D, and two component models that include both slab and 2D 
contributions (e.g. Matthaeus et al. 1990).

There are recent spacecraft measurements of magnetic correlations in the solar wind (see e.g. 
Matthaeus et al. 2005, Dasso et al. 2007). Such measurements are very interesting and important 
since they allow an improved understanding of turbulence. For instance, characteristic length scales 
of turbulence such as the correlation length, the bendover scale, and the dissipation scale can be 
obtained from such observations. Also the investigation of spectral anisotropy by using data from 
different space craft missions such as Wind and ACE is possible. These properties of solar wind 
turbulence are very important for several investigations (heating and damping of the solar wind 
plasma, transport of charged cosmic rays). A further important turbulence property is the
turbulence dynamics (the time dependence of the stochastic magnetic fields). In principle, data sets
from Wind and ACE can also be used to compute dynamical correlation functions to
explore the turbulence dynamics.

In a recent article (Shalchi 2008) magnetic correlation functions were computed analytically. Such 
analytical forms of magnetic correlations complement data analysis results such as Matthaeus et al. 
(2005) and Dasso et al. (2007). Since we expect that future data analysis work will also allow the 
investigation of temporal correlation functions, we explore theoretically (numerically and analytically) 
the forms of these Eulerian correlations. These results can be compared with data analysis results
as soon as they are available.

The organization of the paper is as follows: in section 2 we define and discuss the basic parameters 
which are useful for describing turbulence. Furthermore, we explain the slab, the 2D, and the slab/2D 
compositel model. In section 3 we review different models for the turbulence dynamics. In section 4 
we compute Eulerian correlation functions numerically and analytically. In section 5 the results of this 
article are summarized.
\section{General remarks - setting}
\subsection{The turbulence correlation function}
The key function in turbulence theory is the two-point-two-time correlation tensor. For homogenous turbulence its 
components are
\be
R_{lm} (\vec{x},t) = \left< \delta B_l (\vec{x},t) \delta B_m^{*} (\vec{0},0) \right>.
\label{s1e1}
\ee
The brackets $\left< \dots \right>$ used here denote the ensemble average. It is convenient to introduce the 
correlation tensor in the $\vec{k}-$space. By using the Fourier representation
\be
\delta B_l (\vec{x},t) = \int d^3 k \; \delta B_l (\vec{k},t) e^{i \vec{k} \cdot \vec{x}}
\label{s1e2}
\ee
we find
\bdm
R_{lm} (\vec{x},t) = \int d^3 k \int d^3 k^{'} \left< \delta B_{l} (\vec{k},t) \delta B_m^{*} (\vec{k}^{'},0) \right>
e^{i \vec{k} \cdot \vec{x}}.
\label{s1e3}
\edm
For homogenous turbulence we have
\be
\left< \delta B_{l} (\vec{k},t) \delta B_m^{*} (\vec{k}^{'},0) \right> = P_{lm} (\vec{k},t)  \delta (\vec{k} - \vec{k}^{'})
\label{s1e4}
\ee
with the correlation tensor in the $\vec{k}-$space $P_{lm} (\vec{k},t)$. By assuming the same temporal 
behaviour of all tensor components, we have
\be
P_{lm} (\vec{k}, t) = P_{lm} (\vec{k}) \; \Gamma (\vec{k}, t)
\ee
with the dynamical correlation funtion $\Gamma (\vec{k}, t)$. Eq. (\ref{s1e3}) becomes than
\be
R_{lm} (\vec{x},t) = \int d^3 k \; P_{lm} (\vec{k}) \Gamma (\vec{k}, t) e^{i \vec{k} \cdot (\vec{x})}
\label{s1e5}
\ee
with the magnetostatic tensor $P_{lm} (\vec{k}) = \left< \delta B_{l} (\vec{k}) \delta B_m^{*} (\vec{k}) \right>$.
\subsection{The two-component turbulence model}
In this paragraph we discuss the static tensor $P_{lm} (\vec{k})$ defined in Eq. (\ref{s1e5}). Matthaeus \& 
Smith (1981) have investigated axisymmetric turbulence and derived a general form of $P_{lm} (\vec{k})$ 
for this special case. In our case the symmetry-axis has to be identified with the axis of the uniform mean 
magnetic field $\vec{B}_0 = B_0 \vec{e}_z$. For most applications (e.g. plasma containment devices, 
interplanetary medium) the condition of axisymmetry should be well satisfied. Furthermore, we neglect 
magnetic helicity and we assume that the parallel component of the turbulent fields is zero or negligible 
small ($\delta B_z = 0$). In this case the correlation tensor has the form
\be
P_{lm} (\vec{k}) = A(k_{\parallel},k_{\perp}) \left[ \delta_{lm} - \frac{k_l k_m}{k^2} \right], \quad l,m = x,y
\label{s1e20}
\ee
and $P_{lz}=P_{zm}=0$. The function $A(k_{\parallel},k_{\perp})$ is controlled by two turbulence properties: 
the turbulence geometry and the turbulence wave spectrum. The geometry describes how 
$A(k_{\parallel},k_{\perp})$ depends on the direction of the wave vector $\vec{k}$  with respect to the 
mean field. There are at least three established models for the turbulence geometry:
\begin{enumerate}
\item The slab model: here we assume the form
\be
A^{slab} (k_{\parallel},k_{\perp}) = g^{slab} (k_{\parallel}) \frac{\delta (k_{\perp})}{k_{\perp}}.
\label{s1e21}
\ee
In this model the wave vectors are aligned parallel to the mean field ($\vec{k} \parallel \vec{B}_0$).
\item The 2D model: here we replace $A(k_{\parallel},k_{\perp})$ by
\be
A^{2D} (k_{\parallel},k_{\perp}) = g^{2D} (k_{\perp}) \frac{\delta (k_{\parallel})}{k_{\perp}}.
\label{s1e22}
\ee
In this model the wave vectors are aligned perpendicular to the mean field ($\vec{k} \perp \vec{B}_0$)
and are therefore in a two-dimensional (2D) plane.
\item The slab/2D composite (or two-component) model: In reality the turbulent fields can depend on 
all three coordinates of space. A quasi-three-dimensional model is the so-called slab/2D composite
model, where we assume a superposition of slab and 2D fluctuations:
$\delta B_i^{comp} (x,y,z) = \delta B_i^{slab} (z) + \delta B_i^{2D} (x,y)$. Because of 
$< \delta B_i^{slab} (z) \delta B_j^{*,2D} (x,y) > = 0$, the correlation tensor has the form
\be
P_{lm}^{comp} (\vec{k}) = P_{lm}^{slab} (\vec{k}) + P_{lm}^{2D} (\vec{k}).
\label{s4e2}
\ee
In the composite model the total strength of the fluctuations is $\delta B^2 = \delta B_{slab}^2 + \delta B_{2D}^2$.
The composite model is often used to model solar wind turbulence. It was demonstrated by several
authors (e.g. Bieber et al. 1994, 1996) that $20 \%$ slab / $80 \%$ 2D should be realistic in the solar wind at 1 AU heliocentric 
distance.
\end{enumerate}
The wave spectrum describes the wave number dependence of $A (k_{\parallel},k_{\perp})$. In the
slab model the spectrum is described by the function $g^{slab} (k_{\parallel})$ and in the 2D model
by $g^{2D} (k_{\perp})$.

As demonstrated in Shalchi (2008), the combined correlation functions (defined as $R_{\perp}=R_{xx}+R_{yy}$) for 
pure slab turbulence is given by
\be
R_{\perp}^{slab} (z) = 8 \pi \int_{0}^{\infty} d k_{\parallel} \; g^{slab} (k_{\parallel}) \cos (k_{\parallel} z)
\label{corrslab}
\ee
and the correlation function for pure 2D is
\be
R_{\perp}^{2D} (\rho) = 2 \pi \int_{0}^{\infty} d k_{\perp} \; g^{2D} (k_{\perp}) J_0 (k_{\perp} \rho).
\label{s3e12}
\ee
Here $z$ is the distance parallel with respect to the mean magnetic field and $\rho$ denotes the distance
in the perpendicular direction. To evaluate these formulas we have to specify the two wave spectra 
$g^{slab} (k_{\parallel})$ and $g^{2D} (k_{\perp})$.
\subsection{The turbulence spectrum}
In a cosmic ray propagation study, Bieber et al. (1994) proposed spectra of the form
\bdm
g^{slab} (k_{\parallel}) & = & \frac{C(\nu)}{2 \pi} l_{slab} \delta B_{slab}^2 (1 + k_{\parallel}^2 l_{slab}^2)^{-\nu} \nonumber\\
g^{2D} (k_{\perp}) & = & \frac{2 C(\nu)}{\pi} l_{2D} \delta B_{2D}^2 (1 + k_{\perp}^2 l_{2D}^2)^{-\nu}
\edm
with the inertial range spectral index $2 \nu$, the two bendover length scales $l_{slab}$ and $l_{2D}$, and the 
strength of the slab and the 2D fluctuations $\delta B_{slab}^2$ and $\delta B_{2D}^2$. By requiring normalization 
of the spectra 
\be
\delta B^2 = \delta B_x^2 + \delta B_y^2 + \delta B_z^2 
= \int d^3 k \; \left[ P_{xx} (\vec{k}) + P_{yy} (\vec{k}) + P_{zz} (\vec{k}) \right]
\label{s2e4}
\ee
we find
\be
C(\nu) = \frac{1}{2 \sqrt{\pi}} \frac{\Gamma (\nu)}{\Gamma (\nu-1/2)}.
\label{s2e5}
\ee
By combining these spectra with Eqs. (\ref{corrslab}) and (\ref{s3e12}) the slab correlation function
\be
R_{\perp}^{slab} (z) = 4 C(\nu) \delta B_{slab}^2 l_{slab} 
\int_{0}^{\infty} d k_{\parallel} \; (1+k_{\parallel}^2 l_{slab}^2)^{-\nu} \cos (k_{\parallel} z)
\label{s2e6}
\ee
as well as the 2D correlation function
\be
R_{\perp}^{2D} (\rho) = 4 C (\nu) \delta B_{2D}^2 l_{2D}
\int_{0}^{\infty} d k_{\perp} \; (1+k_{\perp}^2 l_{2D}^2)^{-\nu} J_0 (k_{\perp} \rho)
\label{s3e13}
\ee
can be calculated. In Eq. (\ref{s3e13}) we have used the Bessel function $J_0 (x)$. In Shalchi (2008)
such calculations valid for magnetostatic turbulence are presented.
\subsection{Correlation functions for dynamical turbulence}
For dynamical turbulence the slab and the 2D correlation functions from Eqs. (\ref{corrslab}) and (\ref{s3e12})
become
\bdm
R_{\perp}^{slab} (z) & = & 8 \pi \int_{0}^{\infty} d k_{\parallel} \; g^{slab} (k_{\parallel}) 
\cos (k_{\parallel} z) \Gamma^{slab} (k_{\parallel},t) \nonumber\\
R_{\perp}^{2D} (\rho) & = & 2 \pi \int_{0}^{\infty} d k_{\perp} \; g^{2D} (k_{\perp}) 
J_0 (k_{\perp} \rho) \Gamma^{2D} (k_{\perp},t).
\label{generalcorr}
\edm
For the model spectrum defined in the previous paragraph these formulas become
\bdm
R_{\perp}^{slab} (z,t) & = & 4 C(\nu) \delta B_{slab}^2 l_{slab}
\int_{0}^{\infty} d k_{\parallel} \; (1+k_{\parallel}^2 l_{slab}^2)^{-\nu} \cos (k_{\parallel} z) \; \Gamma^{slab} (k_{\parallel},t) \nonumber\\
R_{\perp}^{2D} (\rho,t) & = & 4 C (\nu) \delta B_{2D}^2 l_{2D}
\int_{0}^{\infty} d k_{\perp} \; (1+k_{\perp}^2 l_{2D}^2)^{-\nu} J_0 (k_{\perp} \rho) \; \Gamma^{2D} (k_{\perp},t).
\label{corrdyn2}
\edm
To evaluate these equations we have to specify the dynamical correlation functions $\Gamma^{slab} (k_{\parallel},t)$
and $\Gamma^{2D} (k_{\perp},t)$ which is done in the next section.
\section{Dynamical turbulence and plasma wave propagation effects}
In the following, we discuss several models for the dynamical correlation function $\Gamma (\vec{k}, t)$. In 
Table \ref{dyntab}, different models for the dynamical correlation function are summarized and compared 
with each other.
\begin{table}[t]
\begin{center}
\begin{tabular}{|l|l|l|}\hline
$ \textnormal{Model}									$ & $ \Gamma^{slab} (k_{\parallel},t)		
$ & $  \Gamma^{2D} (k_{\perp},t)	$ \\ 
\hline\hline
$ \textnormal{Magnetostatic model}						$ & $ 1								
$ & $ 1 $ \\
$ \textnormal{Damping model of dynamical turbulence}	$ & $ e^{-\alpha v_A k_{\parallel} t}		
$ & $ e^{-\alpha v_A k_{\perp} t} $ \\
$ \textnormal{Random sweeping model}					$ & $ e^{-(\alpha v_A k_{\parallel} t)^2}	
$ & $ e^{-(\alpha v_A k_{\perp} t)^2} $ \\
$ \textnormal{Undampled shear Alfv\'en waves}			$ & $ \cos (\pm v_A k_{\parallel} t)		
$ & $ 1 $ \\
$ \textnormal{Undampled fast mode waves}				$ & $ \cos (v_A k_{\parallel} t)			
$ & $ \cos (v_A k_{\perp} t) $ \\
$ \textnormal{NADT model}							$ & $ \cos (\pm v_A k_{\parallel} t) e^{-\gamma_{slab} t}	
$ & $ e^{- \gamma_{2D} t} $ \\
\hline
\end{tabular}
\medskip
\caption{\it Different models for the dynamical correlation function $\Gamma (\vec{k},t)$. Here, $v_A$ is the 
Alfv\'en speed and $\alpha$ is a parameter that allows to adjust the strength of dynamical effects. The 
parameters $\gamma_{slab}$ and $\gamma^{2D}$ of the NADT model are defined in Eq. (\ref{c2s6e3}).}
\label{dyntab}
\end{center}
\end{table}
In the $\vec{k}-$space these models show a very different decorrelation in time. In the Section IV
we compute the dynamical correlation functions in the configuration space for these different models
\subsection{Damping and random sweeping models}
One of the first authors discussing dynamical turbulence were Bieber et al. (1994). In their article,
the authors proposed two models for the dynamical correlation function:
\bdm
\Gamma_{DT} (\vec{k} ,t) & = & e^{- t/t_c(\vec{k})} \quad \textnormal{(damping model of dynamical turbulence)} \nonumber\\
\Gamma_{RS} (\vec{k} ,t) & = & e^{- (t/t_c(\vec{k}))^2} \quad \textnormal{(random sweeping model)}
\label{c2s5e2}
\edm
with the correlation time scale $t_c(\vec{k})$. Bieber et al. (1994) estimated the correlation time as
\be
\frac{1}{t_c(\vec{k})} = \alpha v_A \mid \vec{k} \mid.
\label{c2s5e3}
\ee
Here, $v_A$ is the Alfv\'en speed and $\alpha$ is a parameter which allows to adjust the strength of the dynamical
effects, ranging from $\alpha=0$ (magnetostatic turbulence) to $\alpha=1$ (strongly dynamical turbulence). Bieber et al. (1994)
also suggested that the parameter $\alpha$ could be interpreted as $\delta B / B_0$. In this case, the correlation time scale
$t_c(\vec{k})$ becomes comparable to the eddy turnover time. Also, decorrelation effects related to plasma waves (see, e.g., Schlickeiser
\& Achatz 1993) can be achieved by expressing $\alpha$ through parameters such as the plasma $\beta$ (for a definition see also
Schlickeiser \& Achatz 1993).
The damping model was originally introduced for a particle scattering study in dynamcial turbulence by
Bieber et al. (1994). In this model, the dynamical correlation function has an exponential form, whereas in the
random sweeping model $\Gamma (\vec{k} ,t)$ has a Gaussian form.
\subsection{Plasma wave turbulence}
Another prominent model is the plasma wave model which is discussed in Schlickeiser (2002). In this model, the
dynamical correlation function has the form
\be
\Gamma_{PW} (\vec{k} ,t) = \cos (\omega t) e^{- \gamma t}.
\label{c2s5e4}
\ee
Here, $\omega$ is the plasma wave dispersion relation, whereas $\gamma$ desribes plasma wave damping.
Often, undamped plasma waves are considered, where $\Gamma_{PW} (\vec{k} ,t) = \cos (\omega t)$ and the 
dynamical correlation function is a purely oszillating function. Prominent examples for different plasma waves 
are Shear Alfv\'en waves, where $\omega=\pm v_A k_{\parallel}$, and fast magnetosonic waves, where 
$\omega=v_A k$. 
\subsection{The nonlinear anisotropic dynamical turbulence model}
Recently, an improved dynamical turbulence model, namely the nonlinear anisotropic dynamical turbulence 
(NADT) model, has been proposed by Shalchi et al. (2006). This model takes into account plasma wave 
propagation effects as well as dynamical turbulence effects. The NADT model was formulated for the slab/2D 
composite model, where, in general, we have the two different dynamical correlation functions 
$\Gamma^{slab} (k_{\parallel},t)$ and $\Gamma^{2D} (k_{\perp},t)$, namely
\bdm
\Gamma^{slab} (k_{\parallel},t) & = & \cos (\omega t) e^{- \gamma^{slab} \; t} \nonumber\\
\Gamma^{2D} (k_{\perp},t) & = & e^{- \gamma^{2D} \; t}
\label{c2s6e2}
\edm
with
\bdm
\gamma^{slab} & = & \beta \nonumber\\
\gamma^{2D} & = & \beta \; \left\{
\begin{array}{ccc}
1 & \textnormal{for} & k_{\perp} l_{2D} \leq 1 \\
(k_{\perp} l_{2D})^{2/3} & \textnormal{for} & k_{\perp} l_{2D} \geq 1.
\end{array}
\right.
\label{c2s6e3}
\edm
and with the plasma wave dispersion relation of shear Alfv\'en waves 
\be
\omega = j v_A k_{\parallel} \quad j = \pm 1.
\label{c2s6e4}
\ee
In Eq. (\ref{c2s6e3}) the parameter $\beta$ can be expressed by the strength of the 2D component 
$\delta B_{2D} / B_0$, the 2D bendover scale $l_{2D}$, and the Alfv\'en Speed $v_A$ (see Shalchi 
et al. 2006):
\be
\beta = \sqrt{2} \frac{v_A}{l_{2D}} \frac{\delta B_{2D}}{B_0}.
\label{c2s6e8}
\ee
These forms of the temporal correlation function are discussed in more detail in Shalchi et al. (2006).
They are based on the work of Shebalin (1983), Matthaeus et al. (1990), Tu \& Marsch (1993), 
Oughton et al. (1994), Zhou et al. (2004), Oughton et al. (2006). In the current article we approximate
$\gamma^{2D}$ in Eq. (\ref{c2s6e3}) by
\be
\gamma^{2D} = \beta \left( 1 + k_{\perp} l_{2D} \right)^{2/3}
\ee
for simplicity.

The parameter $j$ in Eq. (\ref{c2s6e4}) is used to track the wave propagation direction ($j=+1$ is used for forward and $j=-1$
for backward to the ambient magnetic field propagating Alfv\'en waves). A lot of studies have addressed the direction of propagation 
of Alfv\'enic turbulence, see, for instance, Bavassano (2003). In general, one would expect that, closer to the sun, most  waves 
should propagate forward and, far away from the sun, the wave intensities should be equal for both directions. Most of the 
observations, which allow conclusions on space plasma and particle propagation properties, have been performed in the solar 
wind at 1 AU heliocentric distance. Thus, we can assume that all waves propagate forward, and we therefore set $j=+1$
in the current article.
\section{Eulerian correlation functions}
To investigate the different models we calculate the (combined) single-point-two-time correlation function defined by
\be
E_{\perp} (t) := R_{\perp} (\vec{x}=0,t) = \left< \delta B_l (0, t) \delta B_m^{*} (0, 0) \right>.
\ee
Since this function is of particular importance in understanding dynamical turbulence effects and interactions
with energetic charged particles this function is also known as Eulerian correlation function $E_{\perp}(t)$.
For the different models discussed in Section III, Eqs. (\ref{corrdyn2}) become
\bdm
E_{\perp}^{slab} (t) & = & 4 C (\nu) \delta B_{slab}^2 \int_{0}^{\infty} d x \; (1 + x^2 )^{-\nu} \nonumber\\
& \times & \left\{
\begin{array}{cc}
1 & \textnormal{magnetostatic model} \\
\cos (\tau x) & \textnormal{Alfv\'en waves} \\
e^{- \alpha \tau x} & \textnormal{damping model} \\
e^{- (\alpha \tau x)^2} & \textnormal{random sweeping model} \\
\cos (\tau x) e^{- \xi \tau} & \textnormal{NADT model}
\end{array}
\right.
\label{c2s5e9}
\edm
and
\bdm
E_{\perp}^{2D} (t) & = & 4 C (\nu) \delta B_{2D}^2 \int_{0}^{\infty} d x \; (1 + x^2 )^{-\nu} \nonumber\\
& \times & \left\{
\begin{array}{cc}
1 & \textnormal{magnetostatic model} \\
1 & \textnormal{Alfv\'en waves} \\
e^{- \alpha \frac{l_{slab}}{l_{2D}} \tau x} & \textnormal{damping model} \\
e^{- (\alpha \frac{l_{slab}}{l_{2D}} \tau x)^2} & \textnormal{random sweeping model} \\
e^{- \xi \tau (1+x)^{2/3}} & \textnormal{NADT model}
\end{array}
\right.
\label{c2s5e10}
\edm
Here we have used the integral transformations $x=k_{\parallel} l_{slab}$ and $x=k_{\perp} l_{2D}$.
Furthermore we used the dimensionless time 
\be
\tau=v_A t / l_{slab}
\ee
and the parameter
\be
\xi = \sqrt{2} \frac{\delta B_{2D}}{B_0} \frac{l_{slab}}{l_{2D}}.
\ee
In the following paragraphs we evaluate Eqs. (\ref{c2s5e9}) and (\ref{c2s5e10}) numerically and analytically.
\subsection{Numerical calculation of Eulerian correlations}
\begin{figure}[t]
\begin{center}
\epsfig{file=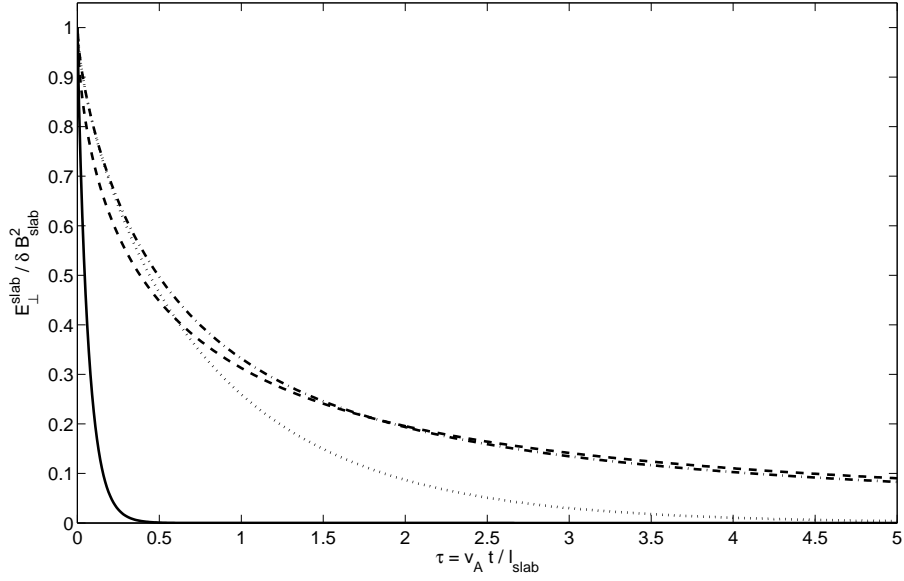, width=400pt}
\end{center}
\caption{The slab correlation function $R_{\perp}^{slab} (0,t) / \delta B_{slab}^2$ as a function of the time 
$\tau=v_A t / l_{slab}$. Shown are the results obtained for the Alfv\'enic plasma wave model (dotted line),
the damping model of dynamical turbulence (dashed line), the random sweeping model (dash-dotted line), 
and the NADT model (solid line).}
\label{corrdynf1}
\end{figure}

\begin{figure}[t]
\begin{center}
\epsfig{file=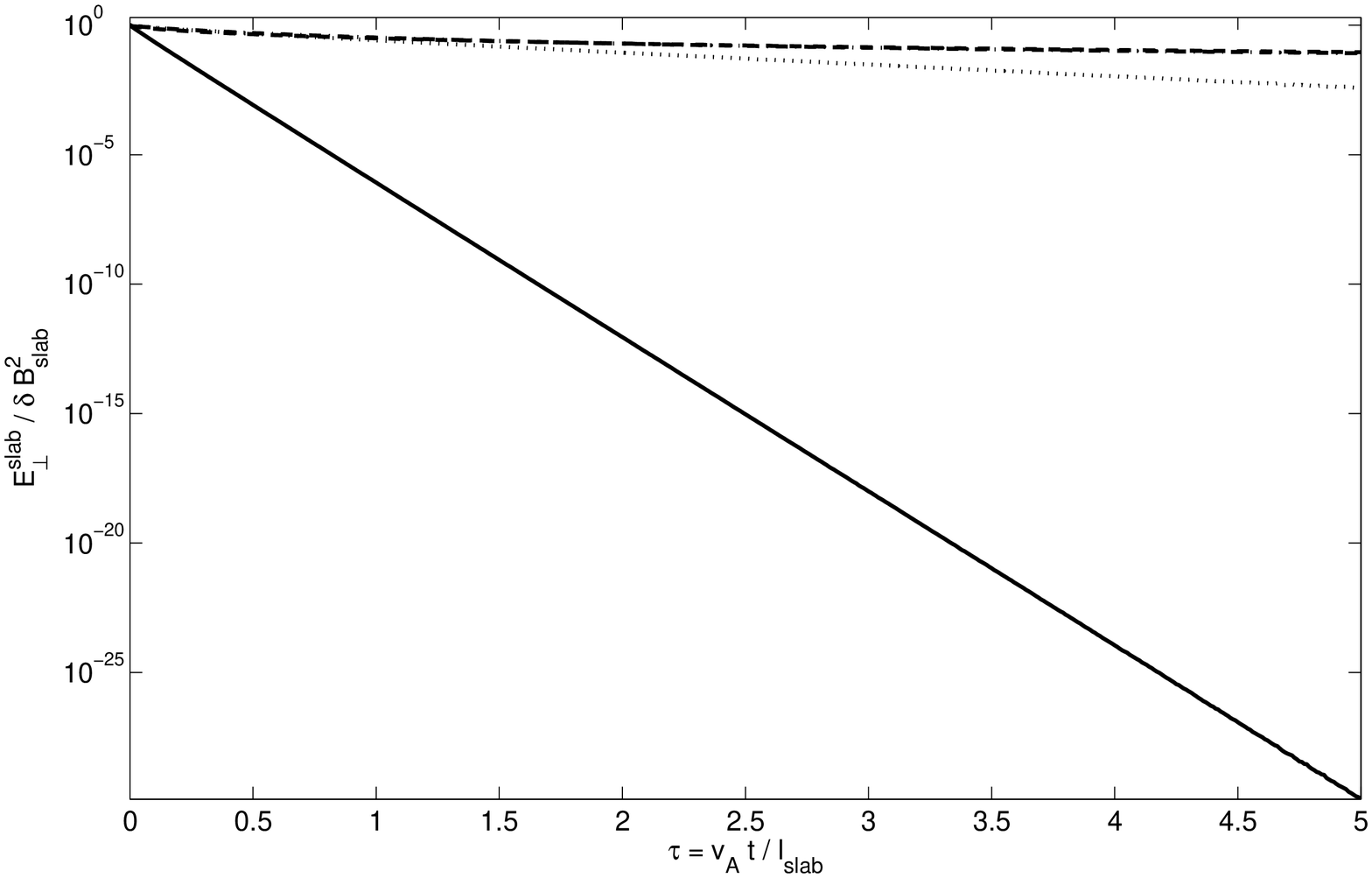, width=400pt}
\end{center}
\caption{Same caption as in Fig. \ref{corrdynf1} but now as semi-log-plot. Clearly we can see that we find an exponential
function for the Eulerian correlation if we employ the Alf\'ven wave model (dotted line) or the NADT model (solid line).}
\label{corrdynf1log}
\end{figure}

\begin{figure}[t]
\begin{center}
\epsfig{file=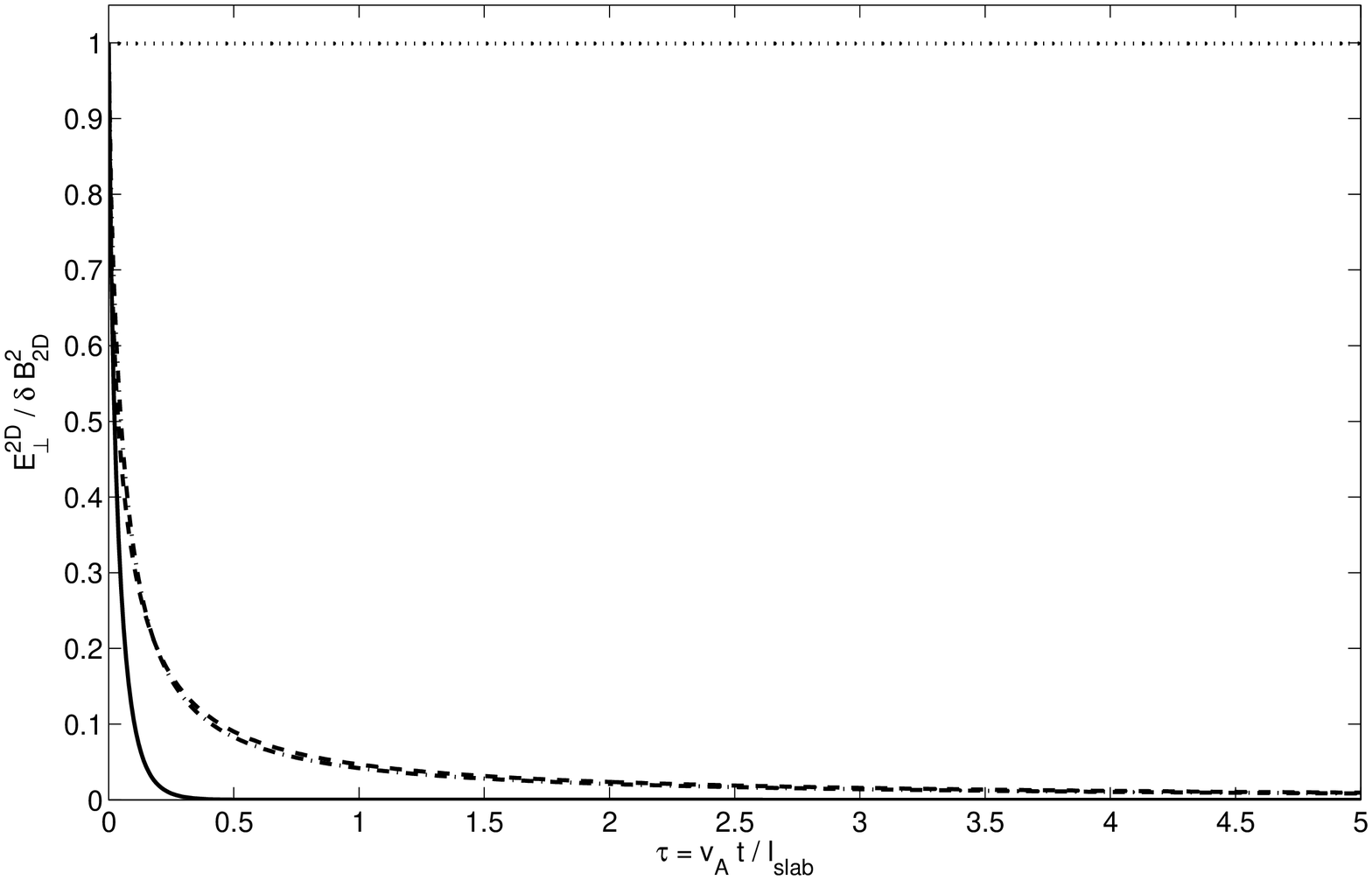, width=400pt}
\end{center}
\caption{The 2D correlation function $R_{\perp}^{2D} (0,t) / \delta B_{2D}^2$ as a function of the time 
$\tau=v_A t / l_{slab}$. Shown are the results obtained for the Alfv\'enic plasma wave model (dotted line),
the damping model of dynamical turbulence (dashed line), the random sweeping model (dash-dotted line), 
and the NADT model (solid line). The result for undampled Alfv\'en waves corresponds to the magnetostatic
model.}
\label{corrdynf2}
\end{figure}

\begin{figure}[t]
\begin{center}
\epsfig{file=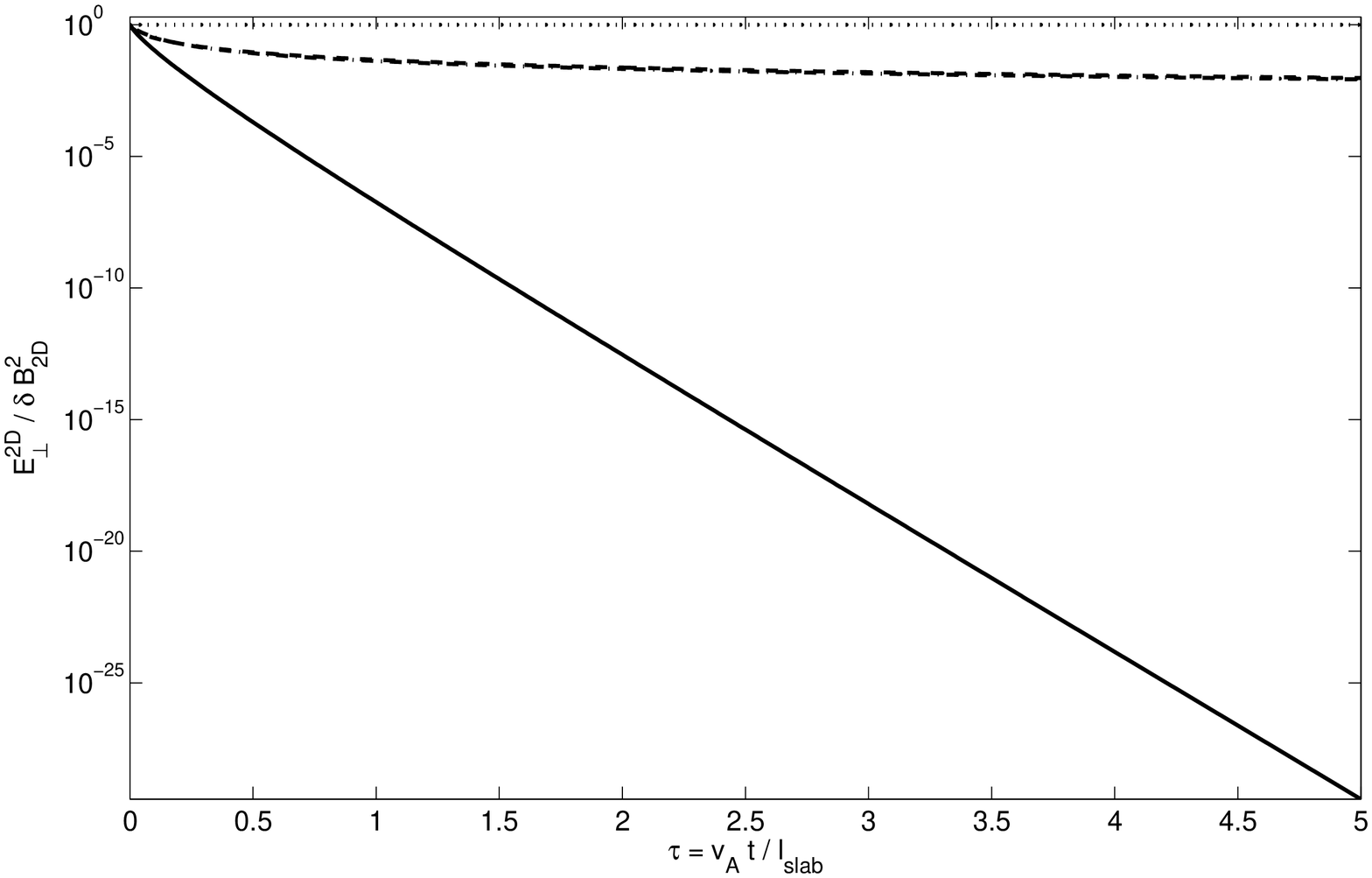, width=400pt}
\end{center}
\caption{Same caption as in Fig. \ref{corrdynf2} but now as semi-log-plot. Clearly we can see that we find an exponential
function for the Eulerian correlation if we employ the NADT model (solid line).}
\label{corrdynf2log}
\end{figure}

In Figs. \ref{corrdynf1} and \ref{corrdynf1log} the results for the slab correlation function and in Figs. \ref{corrdynf2} 
and \ref{corrdynf2log} for the 2D correlation function are shown for the different dynamical turbulence models. To
obtain these figures we have solved the integrals in Eqs. (\ref{c2s5e9}) and (\ref{c2s5e10}) numerically. For
the damping model and the random sweeping model we used $\alpha=1$. Furthermore, we used $l_{2D} = 0.1 l_{slab}$ 
as in previous articles based on the result of laboratory experiments such as Robinson \& Rusbridge (1971). For 
the turbulence spectrum in the inertial range we employ a Kolmogorov (1941) behavior by setting $\nu=5/6$.

As shown in Figs. \ref{corrdynf1} - \ref{corrdynf2log}, the Eulerian correlations obtained for the damping model 
and the random sweeping model are very similar. The results obtained by employing the NADT model are,
however, quite different from the other models.
\subsection{Analytical calculation of Eulerian correlations}
Here we compute analytically the different Eulerian correlations. For magnetostatic (MS) turbulence 
we can use
\be
\int_{0}^{\infty} d x \; (1+x^2)^{-\nu} = \frac{1}{4 C (\nu)}
\ee
and therefore
\bdm
E_{\perp}^{slab,MS} & = & \delta B_{slab}^2 \nonumber\\
E_{\perp}^{2D,MS} & = & \delta B_{2D}^2
\edm
which is the expected result. In the following paragraphs we investigate the different other turbulence models.
\subsubsection{Undampled shear Alfv\'en waves}
In this case we can use (see, e.g., Shalchi 2008)
\be
\int_{0}^{\infty} d x \; (1+x^2)^{-\nu} \cos ( \tau x ) 
= \frac{1}{\Gamma (\nu)} \left( \frac{2}{\tau} \right)^{1/2 - \nu} K_{1/2-\nu} ( \tau)
\label{intalf}
\ee
to derive
\be
E_{\perp}^{slab,Alf} (t) = \frac{4 \delta B_{slab}^2}{\Gamma (\nu-1/2)} \left( \frac{2 l_{slab}}{v_A t} \right)^{1/2 - \nu}
K_{1/2-\nu} \left( \frac{v_A t}{l_{slab}} \right).
\label{eulalf}
\ee
Obviously the characteristic time scale for temporal decorrelation $t_{c}$ is
\be
t_{c}^{slab,Alf} = \frac{l_{slab}}{v_A}.
\ee
Following Shalchi (2008), the Modified Bessel function in Eqs. (\ref{intalf}) and (\ref{eulalf}) can be
approximated for large arguments. We find for times much larger than the temporal decorrelation time
\be
E_{\perp}^{slab,Alf} (t \gg t_{c}) \approx \frac{2 \sqrt{\pi}}{\Gamma (\nu-1/2)} \delta B_{slab}^2 
\left( \frac{2 l_{slab}}{v_A t} \right)^{1- \nu} e^{- v_A t / l_{slab}}.
\label{eulalf2}
\ee
For the special case $\nu=1$ we obtain an exponential function. For the 2D Eulerian correlations we always 
have $E_{\perp}^{2D,Alf} (t) = \delta B_{2D}^2$, since there are no wave propagation effects in the perpendicular
direction for (undamped) Alfv\'enic plasma waves.
\subsubsection{Damping model of dynamical turbulence}
For the damping model of dynamical turbulence (DT) the integrals in Eqs. (\ref{c2s5e9}) and (\ref{c2s5e10})
are difficult to solve. The results can be found in the appendix. As shown there, the characteristic time
scale for the temporal decorrelation is
\bdm
t_{c}^{DT} = \left\{
\begin{array}{cc}
\frac{l_{slab}}{\alpha v_A} & \textnormal{for slab fluctuations} \\
\frac{l_{2D}}{\alpha v_A} & \textnormal{for 2D fluctuations} 
\end{array}
\right.
\label{tcdt}
\edm
For the case $t \gg t_{c}$ corresponding to $a_i \tau \gg 1$ ($i=slab,2D$) with
\bdm
a_{slab} & = & \alpha \nonumber\\
a_{2D} & = & \alpha \frac{l_{slab}}{l_{2D}}
\edm
we can easily compute the Eulerian correlation function approximatelly. For large $a_i \tau$ there is only a 
contribution to the exponential function in Eqs. (\ref{c2s5e9}) and (\ref{c2s5e10}) for $x \rightarrow 0$. Thus, 
we can approximate
\be
\int_{0}^{\infty} d x \; (1+x^2)^{-\nu} e^{- a_i \tau x} \approx \int_{0}^{\infty} d x \; e^{- a_i \tau x}
= \frac{1}{a_i \tau}
\ee
to obtain
\bdm
E_{\perp}^{DT} (t \gg t_{c}) = \frac{4 C(\nu)}{\alpha v_A t} \left\{
\begin{array}{cc}
\delta B_{slab}^2 l_{slab} & \textnormal{for slab fluctuations} \\
\delta B_{2D}^2 l_{2D} & \textnormal{for 2D fluctuations}.
\end{array}
\right.
\edm
For the damping model of dynamical turbulence the Eulerian correlation function tends to zero
with $E_{\perp}^{DT} \sim t^{-1}$.
\subsubsection{Random sweeping model}
For the random sweeping model the analytical results can also be found in the appendix. As shown there
we find the same temporal correlation time scale $t_{c}$ as for the damping model of dynamical
turbulence (see Eq. (\ref{tcdt})). For time scale satisfying $t \gg t_{c}$ we can use
\be
\int_{0}^{\infty} d x \; (1+x^2)^{-\nu} e^{- (a_i \tau x)^2} \approx \int_{0}^{\infty} d x \; e^{- (a_i \tau x)^2}
= \frac{\sqrt{\pi}}{2 a_i \tau}
\ee
to find
\bdm
E_{\perp}^{RS} (t \gg t_{c}) = 2 \sqrt{\pi} C(\nu) \left\{
\begin{array}{cc}
\delta B_{slab}^2 \frac{l_{slab}}{\alpha v_A t} & \textnormal{for slab fluctuations} \\
\delta B_{2D}^2 \frac{l_{2D}}{\alpha v_A t} & \textnormal{for 2D fluctuations}.
\end{array}
\right.
\edm
Obviously the results for the random sweeping model are very similar to the results obtained for
the damping model. This conclusion based on analytical investigations agrees with the
numerical results from Figs. \ref{corrdynf1} - \ref{corrdynf2log}.
\subsubsection{NADT model}
Here we have to distinguish between the slab and the 2D correlation function. For the slab function
we can use the result for Alfv\'en waves with an additional factor $exp (- \xi \tau)$. Therefore,
we find for late times
\be
E_{\perp}^{slab,NADT} (t \gg t_{c}) \approx \frac{2 \sqrt{\pi}}{\Gamma (\nu-1/2)} \delta B_{slab}^2 
\left( \frac{2 l_{slab}}{v_A t} \right)^{1- \nu} e^{- v_A t (1+\xi) / l_{slab}}.
\ee
In this case there are two correlation time scales. The first is associated with the plasma wave (PW)
propagation effects 
\be
t_{c,PW}^{slab,NADT} = \frac{l_{slab}}{v_A}
\ee
and the second is associated with the dynamical turbulence (DT) effects. The latter correlation time is
\be
t_{c,DT}^{NADT} = \frac{l_{slab}}{v_A \xi} = \frac{1}{\sqrt{2}} \frac{B_0}{\delta B_{2D}} \frac{l_{2D}}{v_A}.
\label{nadtscaledt}
\ee
For the 2D fluctuations the situation is more complicated and, thus, the analytical calculations can be
found in the appendix. As demonstrated there the correlation time scale is given by Eq. (\ref{nadtscaledt}).
The behavior of the Eulerian correlation function for late time ($t \gg t_{c,DT}^{NADT}$) is an exponential 
function
\be
E_{\perp}^{2D} \approx 4 C (\nu) \delta B_{2D}^2 e^{- v_A t \xi / l_{slab}}.
\ee
This exponential result agrees with our numerical findings visualized in Figs. \ref{corrdynf2} and \ref{corrdynf2log}.
\section{Summary and conclusion}
In this article we have calculated and discussed Eulerian correlation functions. The motivation for this work are
recent articles from Matthaeus et al. (2005) and Dasso et al. (2007). In these papers it was demonstrated,
that magnetic correlation functions can be obtain from spacecrafts measurements (ACE and Wind).
We expect that from such observations also Eulerian correlations can be obtained. In the current article
we computed analytically and numerically these correlations. These theoretical results are very
useful for a comparison with data obtained from ACE and Wind.

We have employed several standard models for solar wind turbulence dynamics, namely the
(undamped and Alfv\'enic) plasma wave model, the damping model of dynamical turbulence,
the random sweeping model, and the nonlinear anisotropic dynamical turbulence (NADT) model.
All these model are combined with a two-component model and a standard form of the
turbulence wave spectrum. As shown, we find very similar Eulerian correlations for the
damping model and the random sweeping model. Therefore, we expect that in a comparison
between these models and spacecraft data, one can not decide which of these models
is more realistic. The NADT model presented in Shalchi et al. (2006), however, provides
different results in comparison to these previous models. In table \ref{corrtimetab} we have compared the
different correlation time scale derived in this article for the different models.
\begin{table}[t]
\begin{center}
\begin{tabular}{|l|l|l|}\hline
$ \textnormal{Model}									$ & $ t_{c}^{slab}					$ & $  t_{c}^{2D} 				$ \\ 
\hline\hline
$ \textnormal{Magnetostatic model}						$ & $ \infty						$ & $ \infty					$ \\
$ \textnormal{Undampled shear Alfv\'en waves}			$ & $ \frac{l_{slab}}{v_A}			$ & $ \infty					$ \\
$ \textnormal{Damping model of dynamical turbulence}	$ & $ \frac{l_{slab}}{\alpha v_A}		$ & $ \frac{l_{2D}}{\alpha v_A}	$ \\
$ \textnormal{Random sweeping model}					$ & $ \frac{l_{slab}}{\alpha v_A}		$ & $ \frac{l_{2D}}{\alpha v_A}	$ \\
$ \textnormal{NADT model (plasma wave effects)}		$ & $ \frac{l_{slab}}{v_A}			$ &	no effect 				\\
$ \textnormal{NADT model (dyn. turbulence effects)}		$ & $ \frac{1}{\sqrt{2}} \frac{B_0}{\delta B_{2D}} \frac{l_{2D}}{v_A}	$ 
& $ \frac{1}{\sqrt{2}} \frac{B_0}{\delta B_{2D}} \frac{l_{2D}}{v_A} $ \\
\hline
\end{tabular}
\medskip
\caption{\it Comparison for the different correlation time scale found in this article. For the damping model of
dynamical turbulence and the random sweeping model we found the same correlation times. For the NADT model
there are two correlation times, one scale for the plasma wave propagation effects and one scale for the
dynamical turbulence effects.}
\label{corrtimetab}
\end{center}
\end{table}

By comparing the results of this article with spacecraft measurements, we can find out
whether modern models like the NADT model is realistic or not. This would be very
useful for testing our understanding of turbulence. Some results of this article, such as
Eqs. (\ref{generalcorr}) are quite general and can easily be applied for other turbulence 
models (e.g. other wave spectra).
\section*{Acknowledgements}
{\it 
This research was supported by Deutsche Forschungsgemeinschaft (DFG) under the Emmy-Noether program
(grant SH 93/3-1). As a member of the {\it Junges Kolleg} A. Shalchi also aknowledges support by the 
Nordrhein-Westf\"alische Akademie der Wissenschaften.}
\appendix
\section{Exact analytical forms of Eulerian correlation functions}
By using Abramowitz \& Stegun (1974) and Gradshteyn \& Ryzhik (2000) we can compute analytically
the different Eulerian correlation functions defined in Eqs. (\ref{c2s5e9}) and (\ref{c2s5e10}). For the 
plasma wave model the results are given in the main part of the paper.
\subsection{Damping model of dynamical turbulence}
For the damping model of dynamical turbulence we can use
\bdm
\int_{0}^{\infty} d x \; (1+x^2)^{-\nu} e^{-a_i \tau x} & = &\frac{2^{-1/2-\nu} \pi^{3/2} (a_i \tau)^{\nu-1/2}}{\Gamma (\nu)} \nonumber\\
& \times & \left[ 2 J_{\nu-1/2} (a_i \tau) Csc (2 \pi \nu) - J_{1/2-\nu} (a_i \tau) Sec (\pi \nu) + Csc (\pi \nu) H_{1/2-\nu} (a_i \tau) \right].
\edm
Here we used Bessel functions $J_{n} (z)$ and the Struve function $H_{n}(z)$.
By making use of this result we find for the Eulerian correlation function
\bdm
E_{\perp}^{i,DT} (t) & = & 4 C (\nu) \delta B_{i}^2 \frac{2^{-1/2-\nu} \pi^{3/2} (a_i \tau)^{\nu-1/2}}{\Gamma (\nu)} \nonumber\\
& \times & \left[ 2 J_{\nu-1/2} (a_i \tau) Csc (2 \pi \nu) - J_{1/2-\nu} (a_i \tau) Sec (\pi \nu) + Csc (\pi \nu) H_{1/2-\nu} (a_i \tau) \right]
\edm
with $i=slab,2D$ and
\bdm
a_{slab} & = & \alpha \nonumber\\
a_{2D} & = & \alpha \frac{l_{slab}}{l_{2D}}
\edm
for the damping model of dynamical turbulence.
\subsection{Random sweeping model}
For the random sweeping model we can employ
\bdm
\int_{0}^{\infty} d x \; (1+x^2)^{-\nu} e^{-(a_i \tau x)^2}
= \frac{\sqrt{\pi}}{2} U \left( \frac{1}{2},\frac{3}{2} - \nu, a_i^2 \tau^2 \right)
\edm
with the confluent hypergeometric function $U(a,b,z)$. By employing this result we find for the 
Eulerian correlation function
\be
E_{\perp}^{i,RS} (t) = 2 \sqrt{\pi} C (\nu) \delta B_{i}^2 U \left( \frac{1}{2},\frac{3}{2} - \nu, a_i^2 \tau^2 \right)
\ee
for the random sweeping model.
\subsection{NADT model}
For the NADT model we only have to explore the 2D fluctuations (the slab result is trivial and discussed in the
main part of the this paper). Eq. (\ref{c2s5e10}) can be rewritten as
\be
E_{\perp}^{2D} = 4 C (\nu) \delta B_{2D}^2 \left[ \int_{0}^{1} d x \; e^{- \xi \tau}
+ \int_{1}^{\infty} d x \; x^{-2 \nu} e^{- \xi \tau x^{2/3}} \right] 
\ee
The first integral in trivial, the second one can be expressed by an exponential integral function $E_{\mu} (z)$:
\be
E_{\perp}^{2D} = 4 C (\nu) \delta B_{2D}^2 \left[ e^{- \xi \tau} + \frac{3}{2} E_{3 \nu - 1/2} (\xi \tau) \right].
\label{ananadt2d}
\ee
This is the final result for the Eulerian correlation function of the 2D fluctuations. To evaluate this expression
for late times ($\xi \tau \gg 1$) we can approximate the exponential integral function by using
\be
E_{3 \nu - 1/2} (\xi \tau \gg 1) \approx \int_{1}^{\infty} e^{- \xi \tau x} = \frac{e^{- \xi \tau}}{\xi \tau}.
\label{approxEn}
\ee
Here we assumed that the main contribution to the integral comes from the smallest values of $x$, namely
$x \approx 1$. By combining Eq. (\ref{approxEn}) with Eq. (\ref{ananadt2d}) we find approximatelly
\be
E_{\perp}^{2D} \approx 4 C (\nu) \delta B_{2D}^2 e^{- \xi \tau}
\ee
corresponding to an exponential behavior of the Eulerian correlation function. For the correlation time
scale we find $\tau_c = 1/\xi$. A further discussion of these results can be found in the main part of
the text.
{}

\end{document}